\documentclass[pdflatex,sn-mathphys-num]{sn-jnl}

\usepackage{graphicx}%
\usepackage{multirow}%
\usepackage{amsmath,amssymb,amsfonts}%
\usepackage{amsthm}%
\usepackage{mathrsfs}%
\usepackage[title]{appendix}%
\usepackage{xcolor}%
\usepackage{textcomp}%
\usepackage{manyfoot}%
\usepackage{booktabs}%
\usepackage{algorithm}%
\usepackage{algorithmicx}%
\usepackage{algpseudocode}%
\usepackage{listings}%

\theoremstyle{thmstyleone}%
\theoremstyle{thmstyletwo}%

\theoremstyle{thmstylethree}%

\begin{document}

\title[Article Title]{An Interpretable Physics Informed Multi-Stream Deep Learning Architecture for the Discrimination between Earthquake, Quarry Blast and Noise
}


\author*[1]{\fnm{Nishtha} \sur{Srivastava}}\email{N.Srivastava@em.uni-frankfurt.de}

\author[1]{\fnm{Johannes} \sur{Faber}}\email{fais.research@proton.me}
\equalcont{These authors contributed equally to this work.}

\author[2]{\fnm{Dhruv Aditya} \sur{Srivastava}}\email{dhruvaditya88@gmail.com}
\equalcont{These authors contributed equally to this work.}

\affil*[1]{\orgname{Goethe University}, \orgaddress{\street{Campus Riedberg}, \city{Frankfurt am Main}, \postcode{60439}, \state{Hessen}, \country{Germany}}}

\affil[2]{ \orgname{Plant Science and Technology Division,
		College of Agriculture, Food and Natural Resources,		
		University of Missouri}, \orgaddress{\street{Street}, \city{Missouri}, \postcode{10587}, \state{State}, \country{US}}}


\abstract{The reliable discrimination of tectonic earthquakes from anthropogenic quarry blasts and transient noise remains a critical challenge in single-station seismic monitoring. In this study, we introduce a novel Physics Informed Convolutional Recurrent Neural Network (PI-CRNN) that embeds seismological domain knowledge directly into the feature extraction and learning process. We adapt a multi-stream architecture with three parallel encoders: (i) Time Domain – SincNet Encoder, (ii) Multi-Resolution Spectrogram branch, and, (iii) Physics Branch, followed by a fusion and a bidirectional-LSTM module. Evaluated on the Curated Pacific Northwest AI-ready Seismic Dataset, the PI-CRNN achieves an overall classification accuracy of 97.56\% on an independent test set.  It outperforms a standard CRNN baseline, a classical P/S amplitude ratio method, and a Physics-Informed Neural Network (PINN) that enforces physical constraints via the loss function. Furthermore, the model demonstrates perfect precision in noise rejection (100\% Recall). Interpretability analysis using saliency maps confirms that the architecture successfully learns distinct physical signatures, identifying bimodal P- and S-wave arrivals for earthquakes versus singular impulsive onsets for blasts. This work establishes a scalable, transparent framework for AI-driven seismology, proving that architectural inductive bias provides an alternative reliable approach compared to purely data-driven approaches. 

}




\maketitle

\section{Introduction}\label{sec1}

Accurate discrimination between earthquakes, anthropogenic explosions, and transient background noise remains a fundamental challenge in observational seismology. This capability is critical for seismological monitoring agencies to ensure the integrity of seismic catalogs (used for hazard assessment) and for ensuring compliance with international treaties such as the Comprehensive Nuclear-Test-Ban Treaty (CTBT) \cite{CTBTO2024}.
As seismic networks are expanding, with time the volume of data generated has far exceeded the capacity for manual analysis. This necessitates the development of reliable, robust, automated classification systems. 

Classically, the discrimination of earthquakes and blasts is performed using physics-based methods, such as the amplitude ratio \cite{Walter1995} and spectral energy distribution \cite{kim1993discrimination}. These methods are rooted in the fundamental source physics: tectonic earthquakes are shear-source mechanisms that generate strong S-waves, whereas explosions are isotropic, compressional sources that produce dominant P-waves with relatively weak shear energy. While these methods are transparent and provide physical interpretability, they often struggle with low-magnitude events, low signal-to-noise ratio (SNR) and/or events recorded at local and regional distances.

As machine learning advanced, it provided an alternative approach to addressing this problem. For instance, \cite{Kim2020} employed a Support Vector Machine (SVM) to discriminate between earthquakes and explosions, using the amplitudes of the P-wave and S-wave components of seismic signals to construct feature vectors. \cite{Renouard2021} investigated the applicability of Random Forest algorithms to classify events recorded during operational seismic monitoring in the Upper Rhine Graben area. \cite{Elkhouly2025} evaluated multiple machine learning techniques including Decision Trees, K-Nearest Neighbors (KNN), Extreme Gradient Boosting (XGBoost), and Logistic Regression to distinguish between natural earthquakes and nuclear explosions.

While classical machine learning significantly advanced automation, deep learning (DL) fundamentally transformed this landscape by enabling neural networks to autonomously learn and extract hierarchical features directly from raw input data. In earthquake monitoring, these approaches have demonstrated remarkable success across multiple tasks, including event detection and seismic phase picking (e.g., \cite{Li2022EPick, Li2024EarthquakeMonitoring, Li2024SAIPy, Mousavi2020EarthquakeTransformer, QuinterosCartaya2025GNSS}), magnitude estimation (e.g., \cite{Chakraborty2022CREIME, Li2024SAIPy, Mousavi2020MLEMagnitude, QuinterosCartaya2024HRGNSS}), and first-motion polarity estimation (e.g., \cite{Li2024SAIPy, Chakraborty2022PolarCAP}). The rapid progress in deep learning–based seismic discrimination has been strongly supported by the development and curation of large, high-quality benchmark datasets, such as STEAD \cite{Mousavi2019STEAD} and INSTANCE \cite{Michelini2021INSTANCE}, which have enabled standardized training, validation, and comparative evaluation of models.

The earliest application of neural networks to the problem of distinguishing earthquakes from blasts dates back to 1990, when \cite{Dowla1990NeuralSeismic} proposed a simple feed forward network trained on regional spectral data which achieved accuracy above 93\% on their test set. In the late 2010s, researchers revisited this problem using deep learning, for instance \cite{Linville2019DeepLearningDiscrimination} explored the use of convolutional and recurrent neural networks to discriminate explosive and tectonic sources for local distances. \cite{Zhou2025LightweightCNN} used a deep learning model based on convolutional neural networks (CNNs) to classify natural earthquakes and blasts, achieving promising results on a dataset recorded in Hainan. \cite{Saad2024CapsuleCNN} combined a Compact Convolutional Transformer with a Capsule Neural Network to distinguish earthquakes from quarry blasts using data recorded by the Egyptian Seismic Network (ENSN) from January to December 2021. \cite{Lu2025EQTypeNet} developed EQTypeNet, a tri-branch CNN that integrates waveform, spectrogram, and event P/S ratio features for seismic event classification and detection across China. Similarly,\cite{Pragnath2025SeismicClassifier} addressed the classification of three categories of seismic events: quarry blasts, earthquakes, and noise, by developing CNN models trained on labeled waveform data recorded at the SUR station of the Gujarat State Seismic Network between 2007 and 2022.

Despite these successes, purely data-driven deep learning models are often viewed as "black boxes" that lack physical transparency and struggle with generalization when applied to regions outside their training data. To address this Physics-Informed Neural Networks (PINNs) have emerged as a promising solution that embeds physical laws directly into neural network training. Early efforts by \cite{Kong2025PINNSeismic} demonstrated that combining deep learning with physics-based features, such as the P/S wave amplitude ratio, significantly enhanced the generalization performance of models when applied to unseen regions. Further advancements have utilized PINNs to solve more complex geophysical problems related to source characterization. \cite{AlfaroDiaz2025SeismicRecord} introduced a Bayesian framework that integrates seismic waveform features with geospatial context to jointly classify event types and quantify their uncertainties.

In the present study we propose a novel Physics-Informed Convolutional Recurrent Neural Network (PI-CRNN) which integrates the seismological domain knowledge into the neural architecture through three specialized branches: (i) Time Domain SincNet Branch, (ii) Multi-Resolution Spectrogram Branch and (iii) Physics Block. The primary contribution of this study is to demonstrate that a physics-informed approach not only matches the accuracy of deep learning models but significantly enhances the reliability. By bridging the gap between classical signal processing and deep learning, this work provides a transparent and verifiable framework for the next generation of automated seismic monitoring systems.
\begin{figure}[htbp] 
	\centering
	\includegraphics[width=0.9\textwidth]{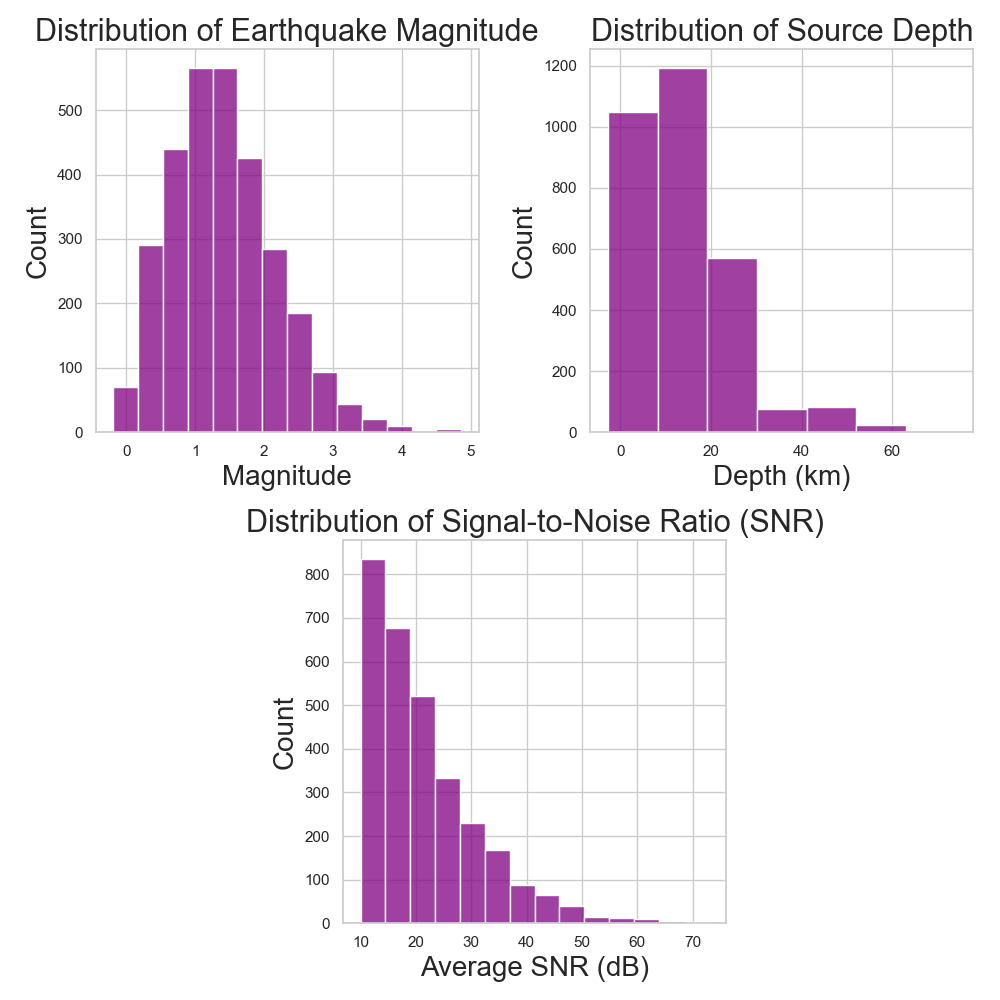} 
	\caption{Statistical distribution of attributes for the earthquake class within the dataset. (Top Left) The magnitude distribution shows majority of events classified as micro-seismicity (Magnitude 1–2). (Top Right) Source depth distribution indicates that most events are shallow crustal earthquakes occurring between 0 and 20 km depth. (Bottom) The distribution of the average Signal-to-Noise Ratio (SNR) shows that the dataset is dominated by lower SNR events (10–30 dB), reflecting the challenging conditions typical of local monitoring environments}
	\label{fig:1} 
\end{figure}

\section{Data and Pre-processing}\label{sec2}
The source data used in this study is the freely available "Curated Pacific Northwest AI-ready Seismic Dataset" \cite{ni2023curated}. This dataset comprises of 190,000 three-component waveforms of more than 65,000 earthquake and explosion events in addition to 9,200 waveforms from 5,600 exotic events accompanied by the event catalog and is freely available via Github.
The dataset derived from \cite{ni2023curated} of the three seismic classes: earthquakes, blasts (explosions), and ambient noise. Event metadata was first filtered to retain traces with a Signal-to-Noise Ratio (SNR) exceeding or equal to 10.0 dB. To construct the final dataset, we targeted a balanced distribution of 3000 training samples and 300 testing samples per class. The sampling rate of raw waveforms is 100 Hz. The original waveforms were of 150-second duration for earthquake and blasts events. The window started 50 seconds prior to the event and ended 100 seconds after the source origin time. For the sake of reduced computation, a fixed 90-second temporal window was extracted for each event, applying a 45-second start offset relative to the record beginning to capture the relevant signal phase. Quality control checks were implemented to automatically reject samples containing 'dead' (all-zero) channels. Finally, each waveform underwent Z-score normalization (subtraction of the mean divided by the standard deviation) to standardize amplitude variance across the dataset. The resulting preprocessed arrays were then saved in NumPy format. The magnitude, depth distribution along with the signal to noise ratio of the selected earthquakes are shown in Figure \ref{fig:1}

The magnitude distribution follows a log-normal distribution consistent with the Gutenberg-Richter law with most of the earthquakes are shallow crustal events (0–20 km) with most magnitude ranging from 1 to 2. Thus the proposed model is primarily trained on microseismic events, which are more frequent but harder to detect in noisy environments.

\section{Method}\label{sec3}
We propose a Physics-Informed Convolutional Recurrent Neural Network (PI-CRNN) to classify seismic events into the three classes: earthquakes, blasts, and noise by integrating seismological features into the deep learning pipeline. Our architecture processes the 3-component seismic data through three parallel feature streams: (1) a learnable SincNet filter bank (2) a multi-resolution spectrogram encoder for time-frequency analysis, and (3) a dedicated physics branch.

\subsection{Model Architecture}\label{subsec1}
\begin{figure}[htbp] 
    \centering
    \includegraphics[width=0.8\textwidth]{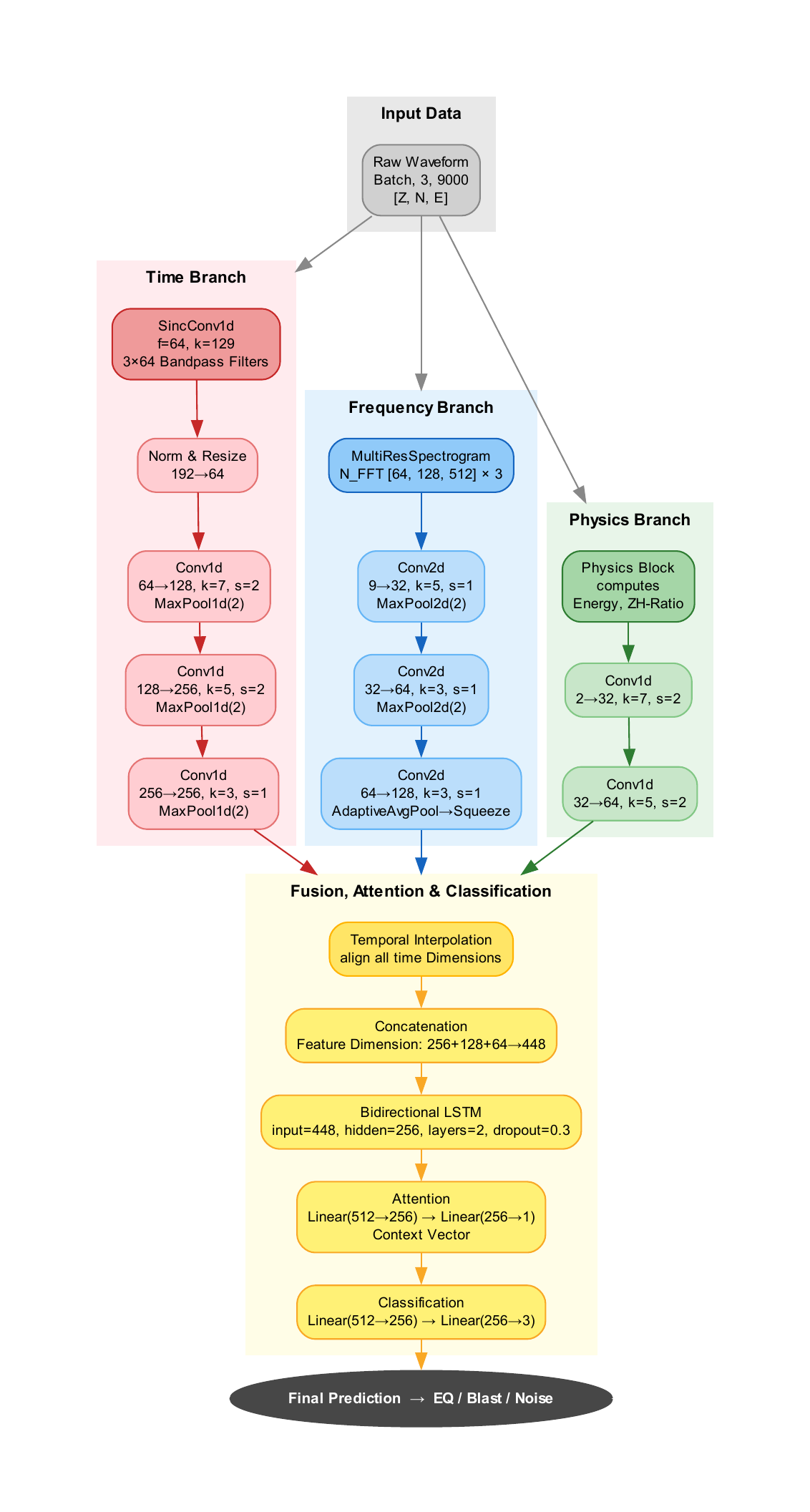} 
    \caption{Model Architecture of PI-CRNN}
    \label{fig:2} 
\end{figure}

The model takes three component input waveforms of 90 seconds. The architecture consists of three parallel encoders followed by a fusion module and a bidirectional-LSTM classifier. The details of the architecture is shown in Figure \ref{fig:2}.
\subsubsection{Stream 1 - Time-Domain SincNet Encoder}\label{subsec1}
To capture specific frequency bands relevant to seismic discrimination we utilize SincNet layers, proposed by \cite{ravanelli2018interpretable} to process raw audio samples, rather than standard convolutional layers. SincNet, based on parametrized sinc functions which implement band-pass filters, encourages the first convolutional layer to discover more meaningful filters. In standard CNNs, the filter weights are randomly initialized and unconstrained, SincNet injects a strong inductive bias by forcing the kernels to approximate distinct band-pass filters.

The SincNet layer is configured with 64 learnable filters applied independently to each of the three input channels, resulting in 64×3=192 output feature maps. This layer is followed by a 1D Convolutional Neural Network (CNN) designed to extract hierarchical temporal features. 
The channel depth is progressively increased and the HardSwish activation \cite{howard2019searching} function is utilized. 

\begin{equation}
\mathrm{HardSwish}(x) = x \cdot \frac{\mathrm{ReLU6}(x + 3)}{6} .
\end{equation}

\subsubsection{Stream 2 - Multi-Resolution Spectrogram}\label{subsec2}
Within this branch, mel-spectrograms are employed to capture both temporal and spectral characteristics of the input signal. Three log-mel spectrograms are computed using the Short-Time Fourier Transform (STFT) with different window lengths and hop sizes.
    \begin{itemize}
        \item A short-window analysis ($n_{\text{fft}} = 64$, hop size $= 16$) with a small number of mel bands (8) to emphasize fine temporal structure,
        \item An intermediate-resolution analysis ($n_{\text{fft}} = 128$, hop size $= 32$) with 16 mel bands, and
        \item A long-window analysis ($n_{\text{fft}} = 512$, hop size $= 64$) with 32 mel bands to capture broader spectral structure over longer time spans.
    \end{itemize}

Due to the different FFT sizes and hop lengths, the resulting spectrograms have mismatched time--frequency dimensions. To enable joint processing, bilinear interpolation is applied to resize the low- and high-resolution spectrograms to match the spatial dimensions of the intermediate-resolution representation. Instead of processing each signal component or each resolution separately, all spectrograms are concatenated along the channel dimension. This produces a 9-channel input tensor (3 resolutions $\times$ 3 components), which is subsequently fed into the convolutional neural network (See Figure \ref{fig:2} for details). 

\subsubsection{Stream 3 -Physics Block}\label{subsec3}

Earthquakes and explosions differ fundamentally in their source mechanism. To leverage this we introduce a non-learnable physics block in the architecture where we calculate following two attributes:

\begin{itemize}
    \item \textbf{Total Instantaneous Energy ($E_{\text{tot}}$):}  
    
    \begin{equation}
    E_{\text{tot}}(t) = Z(t)^2 + N(t)^2 + E(t)^2 .
    \end{equation}

    \item \textbf{Vertical-to-Horizontal Ratio ($R_{ZH}$):}  
    
    \begin{equation}
    R_{ZH}(t) = \frac{Z(t)^2}{N(t)^2 + E(t)^2 + \epsilon} ,
    \end{equation}
    where $\epsilon = 10^{-6}$ is a stability term to prevent division by zero during periods of silence. 
    
\end{itemize}
A high $R_{ZH}$ indicates dominance of compressional waves and a steeper, near-vertically incident wave, characteristic of P-waves from nearby blasts. A low $R_{ZH}$ indicates stronger horizontal energy, characteristic of S-waves from double-couple earthquake sources. In order to prevent gradient explosion while preserving the relative trends, a logarithmic transformation is applied.

\begin{equation}
X_{\text{pol}} = \log\!\left(1 + \left[E_{\text{tot}}, R_{ZH}\right]\right) .
\end{equation}

The addition of unity ensures numerical stability during periods of silence where signal energy approaches zero. These attributes are subsequently processed through convolutional blocks for hierarchical feature learning (see Figure \ref{fig:2}).

\subsubsection{Fusion of Features and Classification}\label{subsec4}
The outputs of the three branches are temporally aligned via linear interpolation and concatenated to form a unified, high-dimensional embedding vector of size
$D_{\text{fusion}} = 256_{\text{(Time)}} + 128_{\text{(Freq)}} + 64_{\text{(Phy)}} = 448$, 
at each time step. This fused sequence is subsequently processed by a two-layer Bidirectional Long Short-Term Memory (Bi-LSTM) network with 256 hidden units, thus allowing the model to capture long-range temporal dependencies. After this a Temporal Attention Mechanism is employed to aggregate the LSTM outputs, thus allowing the network to focus its "attention" on the most informative segments of the signal, such as the impulsive P-wave arrival and the high-energy S-wave phase, while effectively suppressing pre-event noise and post-event scattering. The resulting output is finally passed through dense layers, outputting the final probability distribution over the three target classes (Earthquake, Blast, Noise).

\subsection{Model Training}\label{subsec2}
To mitigate overfitting and improve the model's generalization to unseen seismic events, we employ a probabilistic data augmentation module. During the training phase, each input sample undergoes a series of independent transformations with probability $p = 0.5$. 

\paragraph{Additive Noise Injection.}
To improve robustness under low signal-to-noise ratio (SNR) conditions, an additive white Gaussian noise is introduced as a perturbation. The noise is scaled relative to the signal distribution with a factor of $0.01$:
\begin{equation}
X_{\text{aug}} = X + \varepsilon, \quad \varepsilon \sim \mathcal{N}(0, I) .
\end{equation}

\paragraph{Temporal Shifting.}
To prevent the network from overfitting to a fixed P-wave arrival position (e.g., always expecting the onset at the center of the window), we apply a random cyclic temporal shift $\delta \sim \mathcal{U}(-150, 150)$ samples:
\begin{equation}
X_{\text{aug}}[t] = X[(t - \delta) \bmod L] ,
\end{equation}
where $L$ denotes the length of the waveform. Thus, a random integer between -150 and +150 is chosen and the entire signal is shifted by that many samples.

\paragraph {Mixup Regularization}
In addition to standard data augmentation, following the recent advances in time-series classification \cite{zhu2020seismic}, \cite{choi2024deep}, we employ Mixup regularization \cite{zhang2018mixup} to further smooth the decision boundary and encourage convex behavior in the latent feature space thus,  preventing the 'memorization' of high-frequency site-specific artifacts often associated with quarry blasts.

Given a pair of training samples $(x_i, y_i)$ and $(x_j, y_j)$, we generate a virtual training example $(\tilde{x}, \tilde{y})$ via linear interpolation:
\begin{equation}
    \lambda \sim \text{Beta}(\alpha, \alpha),
\end{equation}
\begin{equation}
    \tilde{x} = \lambda x_i + (1 - \lambda)x_j.
\end{equation}
The training objective is defined as a weighted combination of losses calculated against both original labels:
\begin{equation}
    \mathcal{L}(\theta) = \lambda \mathcal{L}_{CE}(f(\tilde{x}), y_i) + (1 - \lambda) \mathcal{L}_{CE}(f(\tilde{x}), y_j).
\end{equation}
Here, $\alpha = 0.4$ controls the strength of interpolation, and $f(\cdot)$ represents the network parameterized by $\theta$. The base criterion $\mathcal{L}_{CE}$ is implemented as \textit{Cross-Entropy Loss with Label Smoothing} ($\epsilon = 0.1$). By combining Mixup's convex interpolation with Label Smoothing's soft targets, the model is strictly penalized for overconfidence,  improving generalization on ambiguous seismic events.

\paragraph{Training}
The network was trained for 300 epochs with a batch size of 16. The weights corresponding to the highest validation accuracy were saved. Optimization was performed using AdamW \cite{loshchilov2019decoupled} with a weight decay coefficient of 0.02. To accelerate convergence and improve generalization, we employed the One-Cycle Learning Rate policy \cite{smith2019super}, which dynamically schedules both the learning rate and momentum within a single training run. This is a two step process, it begins with a warm-up phase where the learning rate linearly increases to a maximum value (here 0.0001), promoting rapid exploration of the loss landscape and biasing optimization away from sharp minima. This is followed by a cooldown phase where the learning rate decays via cosine annealing to enable fine-grained weight refinement. Momentum is adjusted inversely to the learning rate—reduced during high learning-rate phases to stabilize updates and increased during the cooldown—to further regularize training. Together, these dynamics act as an implicit regularizer, improving robustness to noise and mitigating overfitting.

\section{Results}\label{sec4}

\begin{table}[h]
    \centering
    \caption{Model Classification Performance}
    \label{tab:classification_results}
    \begin{tabular}{lcccc}
        \toprule
        \textbf{Class} & \textbf{Precision} & \textbf{Recall} & \textbf{F1-Score} & \textbf{No. of Samples in Test set} \\
        \midrule
        EQ            & 0.9699 & 0.9667 & 0.9683 & 300 \\
        Blast         & 0.9730 & 0.9600 & 0.9664 & 300 \\
        Noise         & 0.9836 & 1.0000 & 0.9917 & 300 \\
        \midrule
        \textbf{Accuracy} &  &  & \textbf{0.9756} & \textbf{900} \\
        \bottomrule
    \end{tabular}
\end{table}

\begin{figure}[htbp] 
    \centering
    \includegraphics[width=0.8\textwidth]{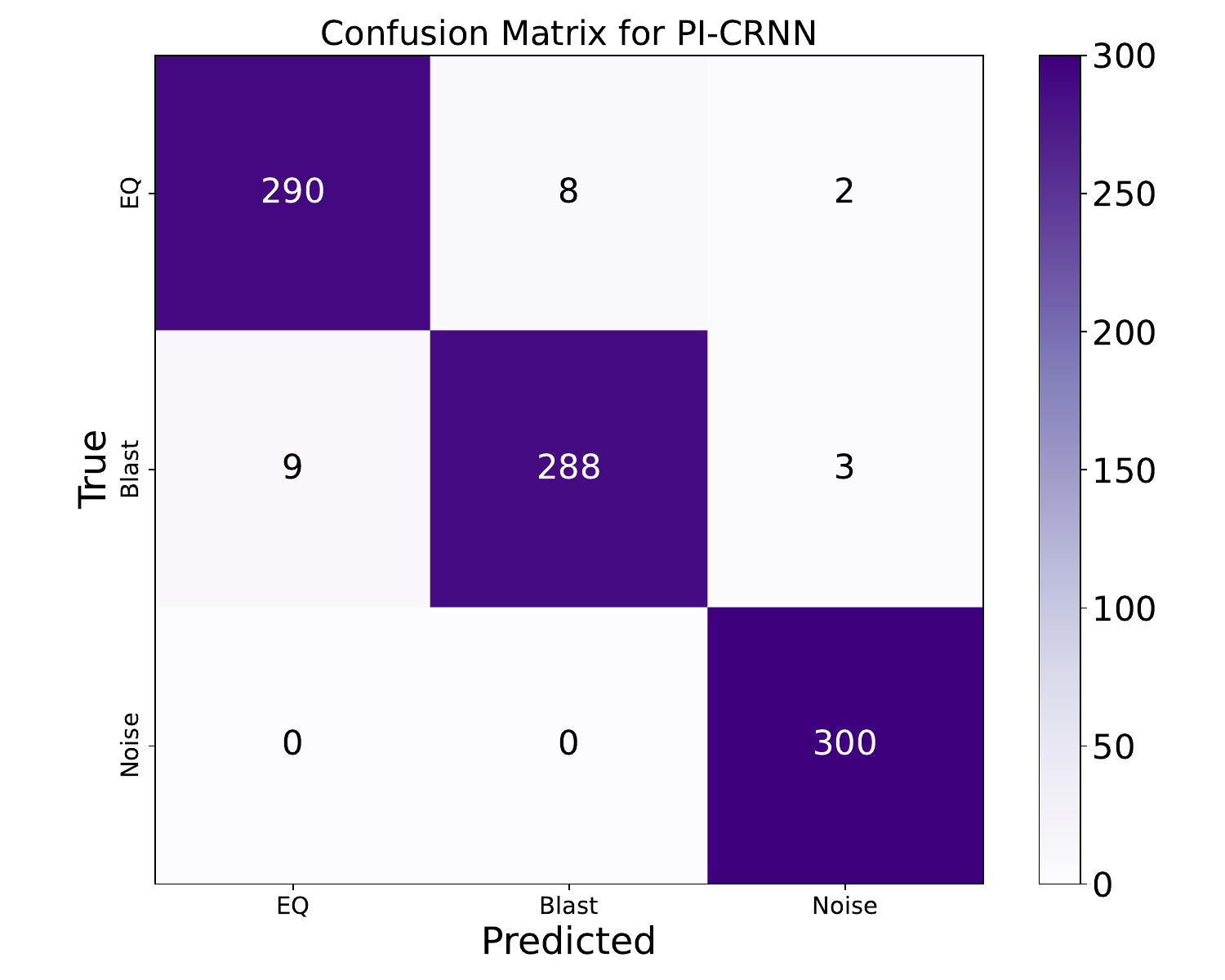} 
    \caption{Confusion Matrix for the proposed PI-CRNN model}
    \label{fig:3} 
\end{figure}

The model achieves high classification performance, as evidenced by the  F1 scores in Table \ref{tab:classification_results} and confusion matrix in Figure \ref{fig:3}. The model successfully distinguishes blasts from local earthquakes thus suggesting that the model learned the impulsive P-wave onset and rapid coda decay (typical behavior of blasts) effectively. Furthermore, the model successfully discriminates all the noise data with near perfect precision. 

Unlike standard ‘black box’ deep learning architectures, PI-CRNN was designed to rely on the seismologically valid features. The Time Domain Branch is designed to process the seismic waveform directly to capture temporal features like phase onsets and polarity. Instead of standard convolution, the branch starts with a SincConv1d layer which forces the filters to be band-pass filters. It uses parametrized Sinc functions to cut out specific frequency bands from the seismic signal. Thus, the network learns to ignore the frequency bands that do not contribute to the correct decision. After this, the signal passes through, a sequence of three Convolutional blocks, transforming input waveform into a compressed sequence of high-level features. Thus, preserving the features which are critical for distinguishing impulsive blasts from slower-onset earthquakes.

While the Time Branch looks at the shape of the wave (the wiggle), Frequency branch looks at the energy distribution across different frequencies over time. We generate three simultaneous views of the input signal via three log-mel spectrograms using the Short-Time Fourier Transform (STFT) with different window lengths and hop sizes, thus, capturing the spectral content decay. Earthquake typically exhibit a broader spectral content with a longer, scattered coda due to complex path effects in the crust. On the other hand, blasts show high-frequency energy concentrated at the onset with faster attenuation. These stacked images are passed through a 2D Convolutional Neural Network (CNN). 

The third physics branch explicitly calculates physical attributes related to the ground motion. This blocks forces the model to calculate two specific seismological variables, total instantaneous energy and Vertical-to-Horizontal Ratio. Thus this branch attempts to introduce the fundamental difference in source mechanisms of earthquakes and blasts.

The positive results of this architecture highlights the value of inductive bias in scientific machine learning. By forcing the network to process data through attributes familiar to seismology (filtering, spectrograms, polarization), we achieved high accuracy with a relatively lightweight model. This system is well-suited for deployment as a prototype automated trigger or classifier in regional seismic networks, potentially reducing the manual workload of analysts by filtering out non-tectonic events with high confidence.
The training and validation loss curves show a steady, monotonic decrease, stabilizing after approximately 150 epochs. The close alignment between training and validation loss indicates that the Mixup regularization and Seismic Augmentation (signal shifting, noise injection) in conjunction with dropout layers effectively prevented overfitting.

\begin{figure}[htbp] 
    \centering
    \includegraphics[width=1.0\textwidth]{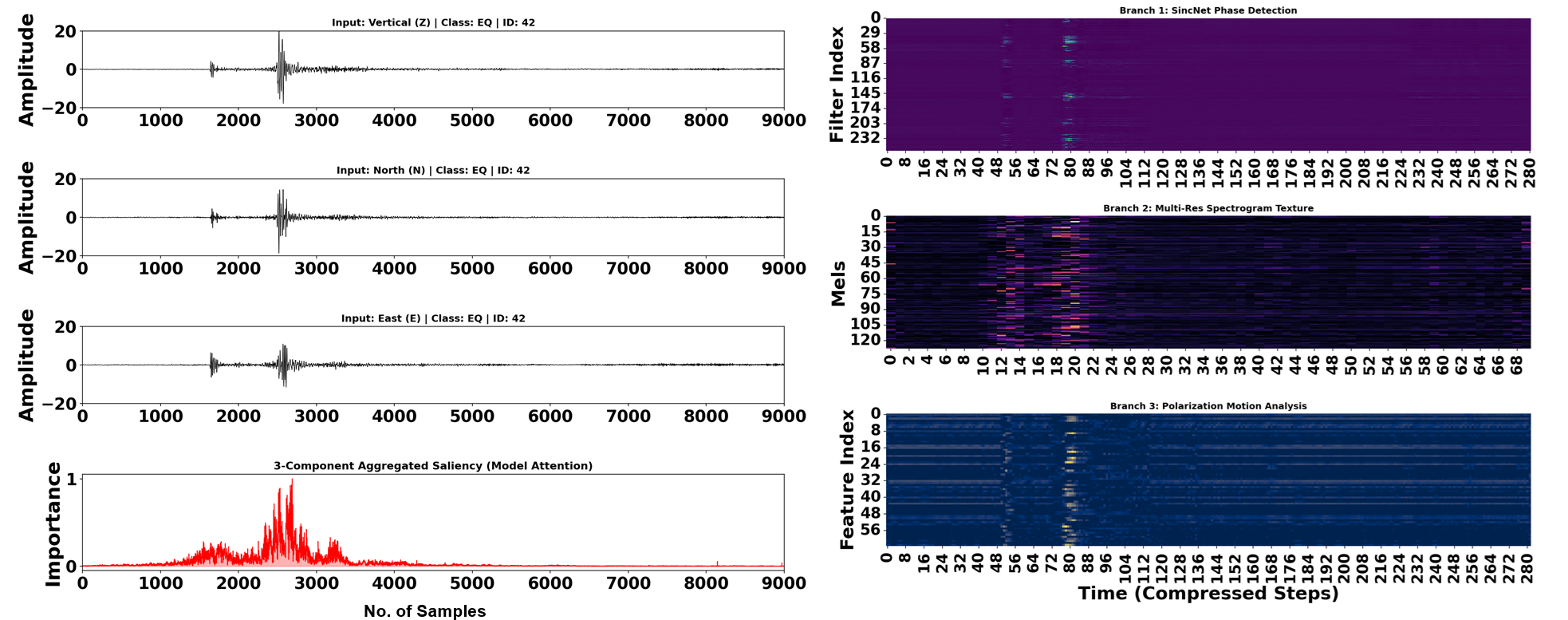} 
    \caption{Tectonic Earthquake: The saliency map exhibits a clear bimodal structure.
    	The first peak aligns with the P-wave onset, followed by a dominant peak corresponding
    	to the S-wave arrival. Consistently, the feature map from the time-domain branch
    	shows a distinct double-activation pattern, indicating that the network has successfully
    	learned to temporally separate the P and S phases.}
    \label{fig:4} 
\end{figure}

\begin{figure}[htbp] 
    \centering
    \includegraphics[width=1.0\textwidth]{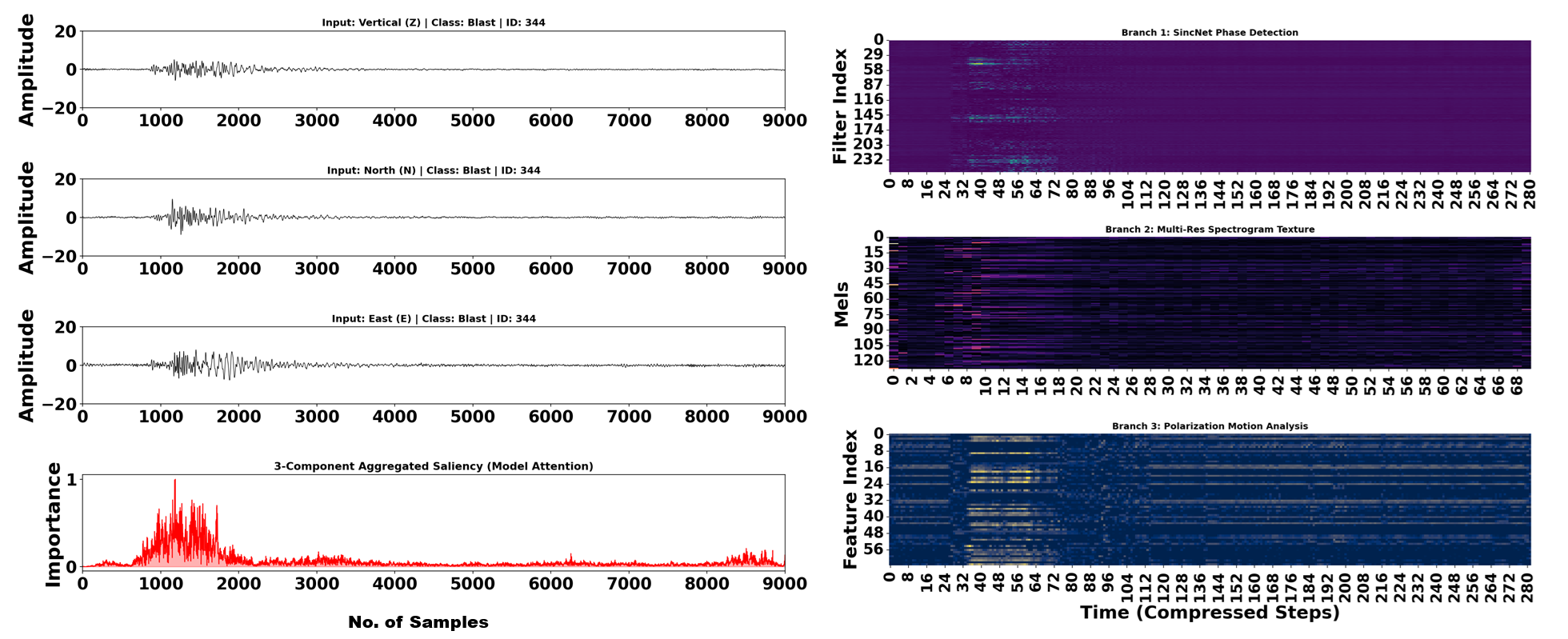} 
    \caption{Blast: The saliency map is characterized by a single
    	dominant peak. The corresponding frequency-domain features display vertically broad
    	bands rather than the two distinct bands observed in the earthquake case, suggesting
    	that the model captures an instantaneous, high-frequency energy release. Likewise,
    	the time-domain feature map exhibits a single, sharp vertical activation.}
    \label{fig:5} 
\end{figure}

\begin{figure}[htbp] 
    \centering
    \includegraphics[width=1.0\textwidth]{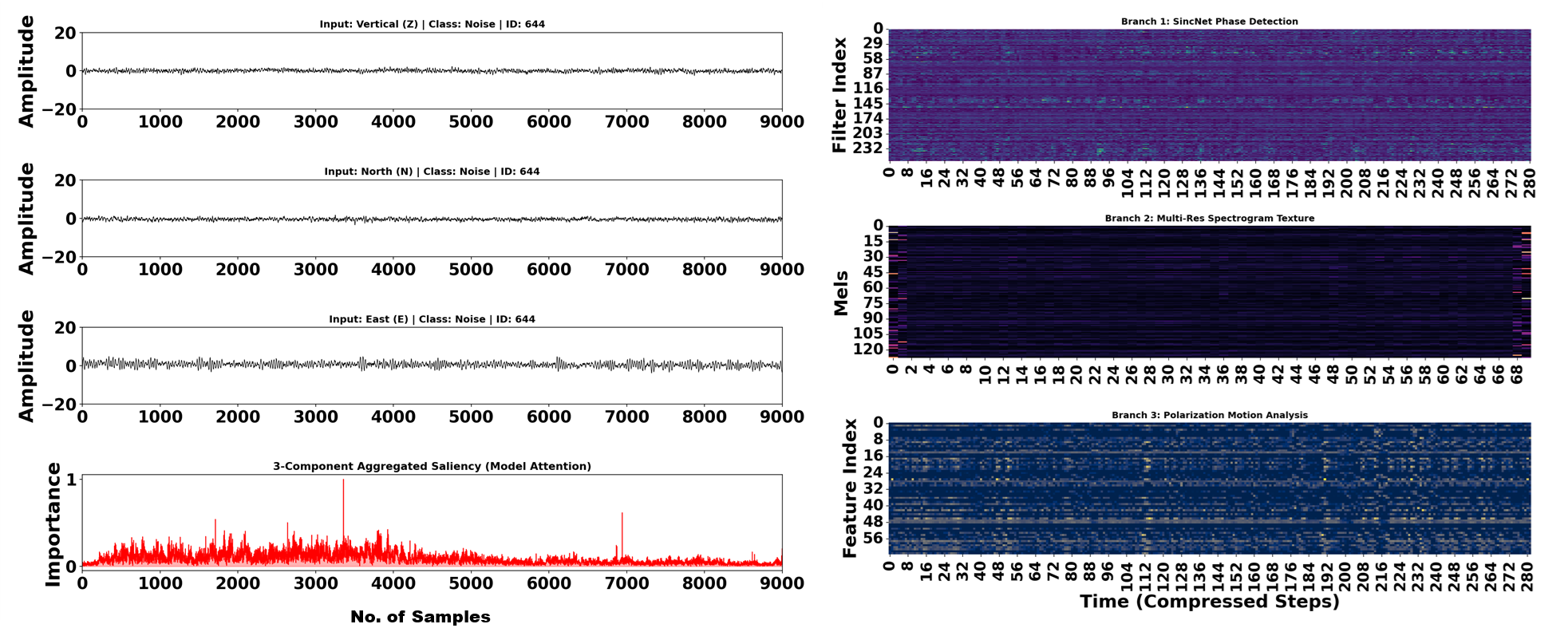} 
    \caption{Noise: The saliency map demonstrates a lack of coherent structure,
    	with only low-amplitude fluctuations and no clear focal region. All three feature maps
    	are more sparsely activated compared to the other two
    	cases, indicating the absence of distinctive, event-like seismological signatures}
    \label{fig:6} 
\end{figure}

To verify that the proposed model is not merely memorizing the training data but is instead learning physically meaningful seismological features that enable high-fidelity classification, we visualize activation heat maps and gradient-based saliency maps for all three classes. Representative examples for each class are shown in Figures \ref{fig:4}, \ref{fig:5}, and \ref{fig:6}.

For the earthquake example Figure \ref{fig:4}, the saliency map exhibits a clear bimodal structure. The first peak aligns with the P-wave onset, followed by a dominant peak corresponding to the S-wave arrival. Consistently, the feature map from the time-domain branch shows a distinct double-activation pattern, indicating that the network has successfully learned to temporally separate the P and S phases.
\begin{figure}[htbp] 
    \centering
    \includegraphics[width=0.8\textwidth]{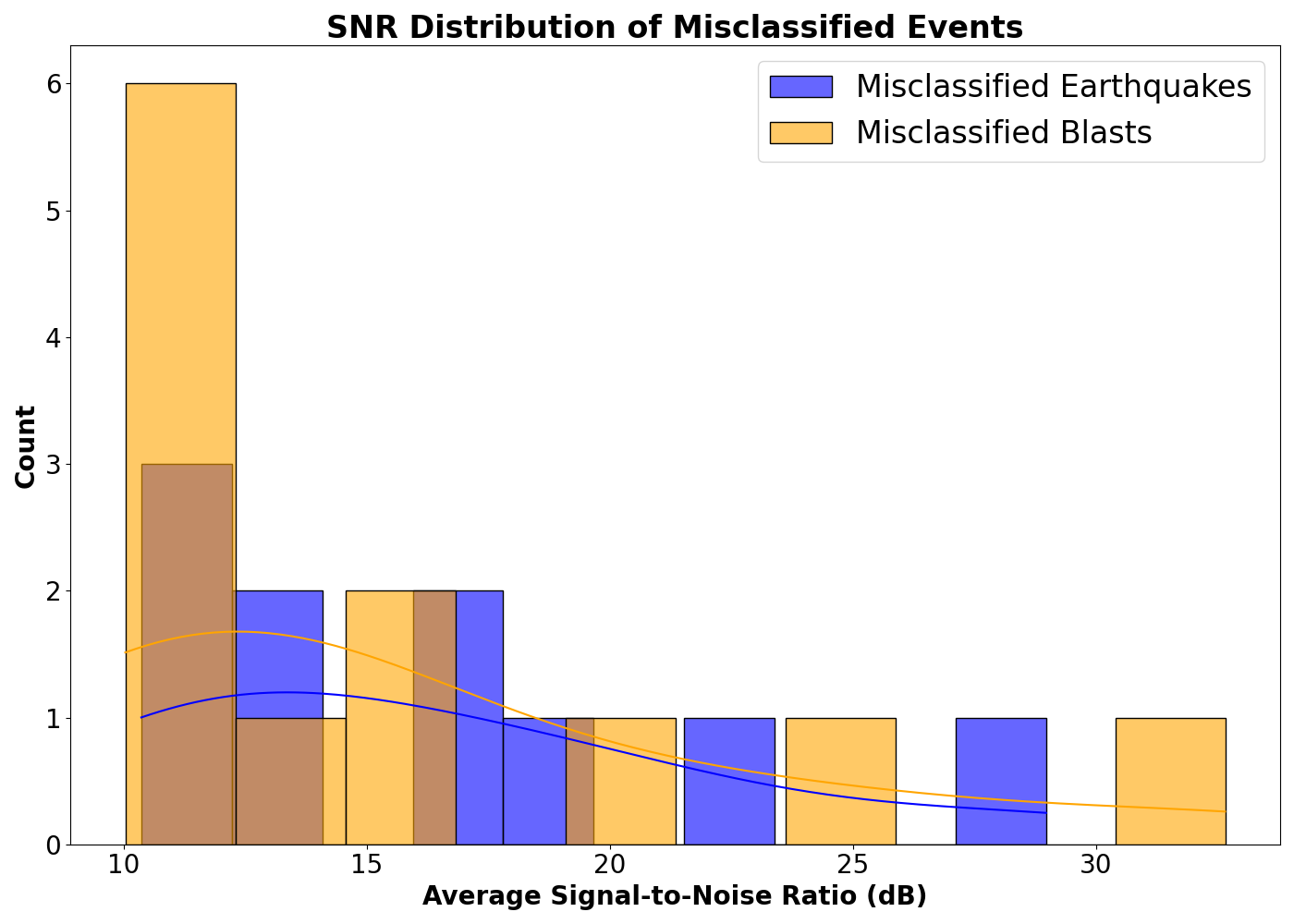} 
    \caption{Distribution of average Signal-to-Noise Ratio (SNR) for misclassified seismic events. The histograms compare the SNR profiles of (a) Earthquakes misclassified as Blasts or Noise (blue) and (b) Blasts misclassified as Earthquakes or Noise (orange). The significant overlap in the 10–15 dB range indicates that low-SNR events are prone to ambiguity.}
    \label{fig:7} 
\end{figure}

\begin{figure}[htbp] 
    \centering
    \includegraphics[width=0.6\textwidth]{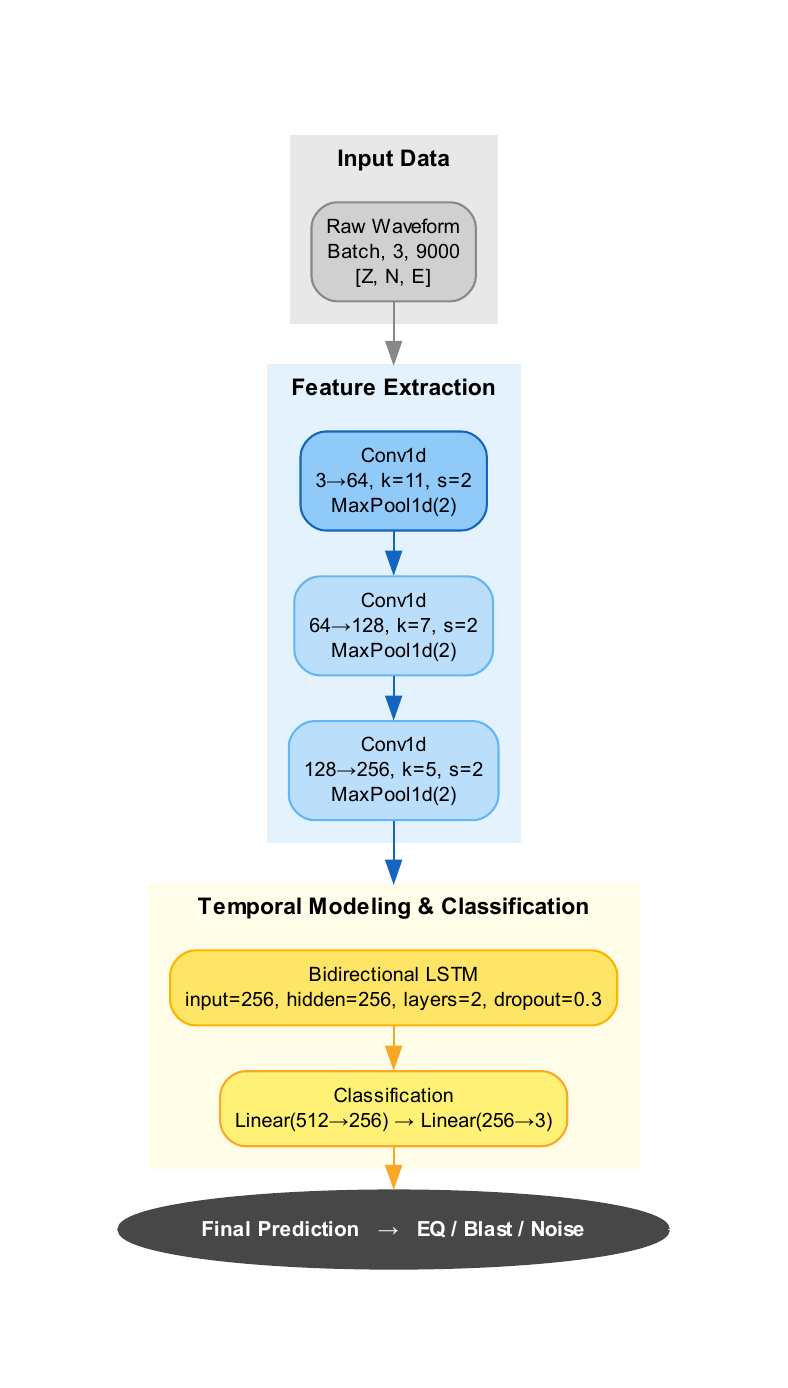} 
    \caption{Baseline Architecture.}
    \label{fig:8} 
\end{figure}

\begin{figure}[htbp] 
    \centering
    \includegraphics[width=0.99\textwidth]{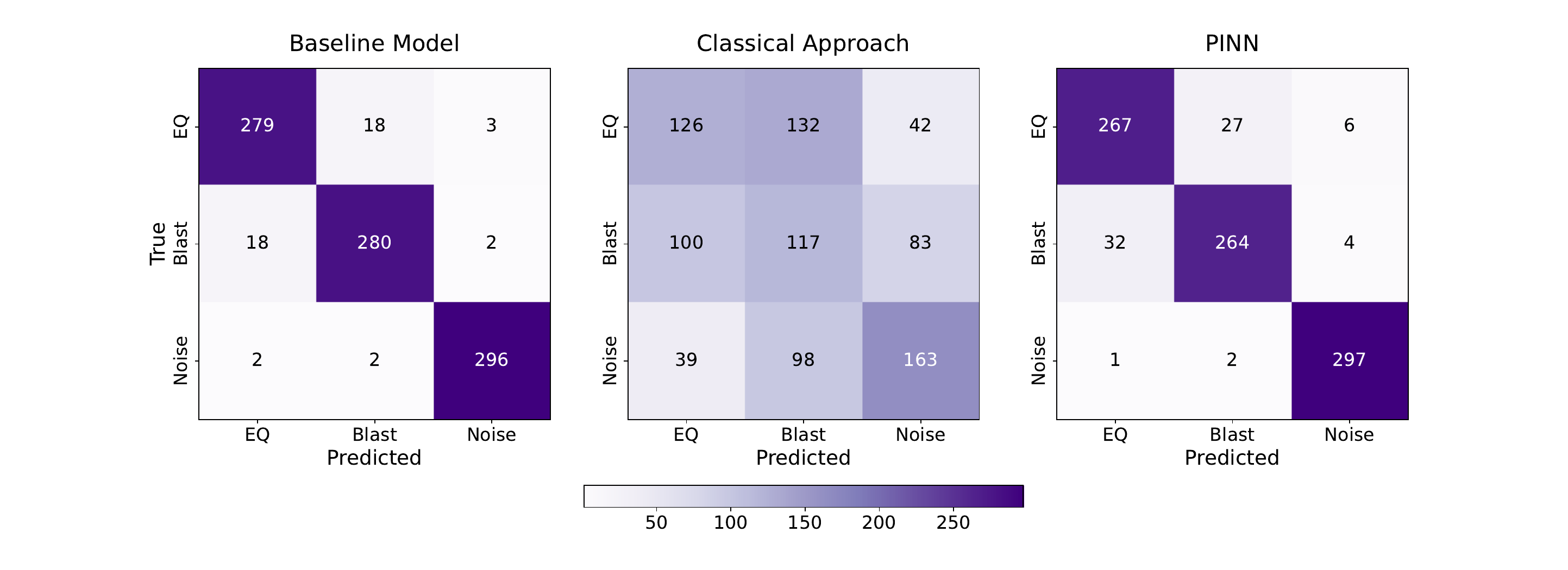} 
    \caption{Confusion Matrix for the comparison models.}
    \label{fig:9} 
\end{figure}

In contrast, the saliency map for the blast example Figure \ref{fig:5} is characterized by a single dominant peak. The corresponding frequency-domain features display vertically broad bands rather than the two distinct bands observed in the earthquake case, suggesting that the model captures an instantaneous, high-frequency energy release. Likewise, the time-domain feature map exhibits a single, sharp vertical activation.

Finally, the saliency map shown in Figure \ref{fig:6} demonstrates a lack of coherent structure, with only low-amplitude fluctuations and no clear focal region. All three feature maps for this class appear darker and more sparsely activated compared to the other two cases, indicating the absence of distinctive, event-like seismological signatures.

\section{Discussion}\label{sec5}
The integration of domain-specific knowledge branches into the architecture provides a robust, physics-guided framework for classifying seismic signals. This is evidenced by the bimodal distribution observed in the saliency map of an earthquake sample, which contrasts sharply with the saliency map of a blast sample, where only a single peak is present. Furthermore, activations in the feature maps confirm that each branch captures a distinct attribute of the seismic trace. These results demonstrate that the model does not merely respond to signal amplitude; rather, it evaluates the temporal and spectral relationships within the waveform to distinguish tectonic events from anthropogenic sources.

A detailed analysis of the misclassified events reveals that the model primarily confuses between earthquakes and blasts. Most earthquake misclassifications correspond to low-magnitude, shallow events with low signal-to-noise ratios (SNR), likely because such near-source earthquakes exhibit impulsive P-wave onsets that resemble those of blasts. Figure \ref{fig:7} presents the SNR distribution for the misclassified events. Notably, both misclassified classes,  \textit{i.e.}, earthquakes and blasts, show substantial overlap in the 10–15 dB range, indicating that low-to-moderate SNR signals are particularly susceptible to confusion. This overlap suggests that distinctive high-frequency characteristics, such as impulsive P-wave onsets, are obscured by background noise in this SNR regime. Nevertheless, a small number of high-SNR outliers are also observed.

\paragraph{Comparison with the non-physics baseline model}

To assess the impact of incorporating domain-specific physical constraints into the architecture, we developed a standard baseline model, shown in Figure \ref{fig:8}. This baseline employs a Convolutional Recurrent Neural Network (CRNN) composed of three one-dimensional convolutional layers to capture temporal patterns, followed by two bidirectional LSTM layers that enable the model to learn long-range sequential dependencies. While this architecture achieves a test-set accuracy of 95\%, it exhibits a substantially higher false-positive rate. Moreover, the baseline lacks the interpretability offered by the physics-informed branches.

The proposed physics-informed model demonstrates performance improvement over the baseline architecture, increasing overall classification accuracy from 95.00\% to 97.56\%. Moreover, the physics-informed approach exhibits comparatively better capability in discriminating between tectonic earthquakes (EQ) and anthropogenic explosions (Blast). While the baseline model struggled with this distinction (Figure \ref{fig:9}, misclassifying 18 earthquakes as explosions and 18 explosions as earthquakes, the proposed model reduced these confusions by over 50\% (8 and 9 misclassifications, respectively). Furthermore, the physics-informed model achieved greater robustness against background noise, attaining a perfect recall of 1.00 for the Noise class in the test set, highlighting its potential in reducing false alarm rates. 

\paragraph{Comparison with the classical approach}
We also compared the performance of the proposed approach with the classical methods by employing a classical STA/LTA trigger (STA=0.5s, LTA=10s, Threshold=3.5) following standard practices for local microseismic monitoring \cite{trnkoczy2012stalta}. Source discrimination was performed using the Shear wave and P Wave amplitude ratio, with a conservative decision threshold of 0.2 to account for S-wave generation via near-surface scattering in quarry blasts \cite{kim1993discrimination}.
Figure \ref{fig:9} displays the confusion matrix derived on the test dataset comprising Earthquake, Blast, and Noise events. The classical method demonstrated limited discriminative capability, achieving an overall accuracy of only 45\%. Most notably, the classical estimator exhibited a severe bias in distinguishing Earthquakes from Blasts; it correctly identified only 126 EQ events while misclassifying a larger portion (132 events) as Blasts. Furthermore, the method struggled to isolate background noise, frequently misidentifying valid signals as noise and vice versa, resulting in high false-positive rates across all classes. This sharp contrast in performance underscores our proposed model's ability to learn high-dimensional, non-linear representations of seismic waveforms, effectively disentangling signal characteristics that remain indistinguishable to heuristic, ratio-based classifiers.

\paragraph{Comparison with the model where physics was introduced in loss function}
A further comparison was made with a Physics-Informed Neural Network (PINN) approach, where physical constraints were applied as a penalty in the loss function. The architecture is based on \cite{Kong2025PINNSeismic} 
which utilized a compact 3-layer Convolutional Neural Network (CNN) optimized for short, 2000-sample windows. We modified the model further adopting a 5-layer structure capable of processing 9000-sample sequences. Other parameters were kept same.  The overall accuracy of this approach on the test set was 92\%, additionally the model struggled in differentiating between earthquakes and blasts (Figure \ref{fig:9}). The results suggest a performance benefit when integrating Physics-Informed Neural Networks (PINNs) directly into the model architecture rather than applying them only through the loss function. While physics-based loss functions provide useful guidance, incorporating physics into the architecture may offer a more consistent way to capture subtle physical patterns in complex seismic data.

\section{Conclusion and Future Work}\label{sec6}
To bridge the gap between deep learning and seismological interpretability, we propose a Physics-Informed Convolutional Recurrent Neural Network (PI-CRNN) for automatic seismic event classification. By integrating domain-specific, physics-guided architectural branches, the proposed model moves beyond a conventional “black-box” approach toward physically grounded decision-making. Saliency and activation maps demonstrate that the network learns discriminative seismological signatures characteristic of earthquakes and explosions. Compared to baseline architecture, PI-CRNN achieves better classification performance Furthermore, this work introduces a transparency in the application of AI in seismology, providing a system that not only detects and classifies events but also offers the physical reasoning behind it. Future work will extend the proposed architecture to multi-station data, which is expected to further reduce false positives. Additionally, in subsequent versions, we plan to incorporate transfer learning to enable regional fine-tuning and improve generalization, as well as self-supervised learning on continuous seismic data streams.

\bmhead{Acknowledgements}

The research has made extensive use of PyTorch \cite{pyTorch2019}, numpy \cite{Numpy2020}, and matplotlib \cite{Matplotlib2007}.

\section*{Data and Code availability}

The datasets supporting the findings of this study are available at the following GitHub repository: https://github.com/niyiyu/PNW-ML/tree/main
.

The model codes presented in this work are available at: https://github.com/srivastavaresearchgroup/PICRNN
.

Code repository is currently maintained in a private state for the purposes of peer review and will be made publicly accessible upon acceptance of the manuscript.

\bmhead{ Author contribution}
Conceptualization: NS; Data Analysis and preparation: NS and DAS; Formal Analysis: NS, Visualization: NS, JF; Methodology: NS, JF, DAS; Original Draft writing: NS; Reviewing and Editing: NS, JF, DAS



\bibliography{sn-bibliography}

\end{document}